\documentclass[twocolumn,aps, prl,superscriptaddress]{revtex4}

\usepackage[colorlinks=true, urlcolor=blue, pdfborder={0 0 0}]{hyperref}

\usepackage{graphicx}
\usepackage{natbib}
\usepackage{hyperref}
\usepackage{amsmath}
\usepackage{mathrsfs}
\usepackage{amsfonts}
\usepackage{color}
\usepackage{bm} 
\usepackage{subfigure}
\usepackage{epsfig} 
\usepackage{epstopdf}
\usepackage{float} 
\usepackage[percent]{overpic}
\usepackage{comment}
\usepackage{pstricks}
\usepackage{filecontents}
\usepackage{enumitem}  

\usepackage{zref-xr}
\usepackage[utf8]{inputenc}

\begin{document}

\title{\bf Entangled Quantum Dynamics of Many-Body Systems using Bohmian Trajectories}

\author{Tarek A. Elsayed}

\thanks{ Correspondence should be addressed to Tarek A. Elsayed}
\email{tarek.ahmed.elsayed@gmail.com}
\address{Zewail City of Science and Technology,6th of October City, Giza 12578, Egypt}

\address{Department of Physics and Astronomy, Aarhus University, 8000 Aarhus C, Denmark}

\author{  Klaus Mølmer }
\address{Department of Physics and Astronomy, Aarhus University, 8000 Aarhus C, Denmark}

\author{  Lars Bojer Madsen }
\address{Department of Physics and Astronomy, Aarhus University, 8000 Aarhus C, Denmark}

\date{\today}


\begin{abstract}

Bohmian mechanics is an interpretation of quantum mechanics that  describes the motion of quantum particles with an ensemble of deterministic trajectories. Several attempts have been made to utilize Bohmian trajectories as a computational tool to simulate quantum systems consisting of many particles, a very demanding computational task. In this paper, we present a novel ab-initio approach to solve the many-body problem for bosonic systems by evolving a system of one-particle wavefunctions representing pilot waves that guide the Bohmian trajectories of the quantum particles. In this approach, quantum entanglement effects arise due to the interactions between different configurations of Bohmian particles evolving simultaneously. The method is used to study the breathing dynamics and ground state properties in a system of  interacting bosons.

\end{abstract}
\maketitle

\section*{Introduction}
Numerical simulation of  the quantum  dynamics of many-body systems is plagued by the dimension of the Hilbert space which increases exponentially with the number of particles. Much of the progress in theoretical condensed matter, atomic and molecular physics in the past few decades has been achieved by finding new ways to circumvent  this problem. Some of the most powerful approaches are density functional theory  \cite{kohn1965}, quantum Monte Carlo  \cite{foulkes2001},  density matrix renormalization group  \cite{white1992} and the multi-configuration time-dependent Hartree method \cite{beck2000}. Recently,  nonconventional approaches based on machine learning \cite{Carleo2017},  Bohmian mechanics \cite{wyatt2006,lopreore1999,benseny2014,hall2014} and wavelet transforms \cite{poirier2003}  have been proposed as well.  Several methods aim to reduce the complexity of the numerical simulation of many-body systems by resorting to low dimensional objects such as density functions \cite{kohn1965,parr1980} and natural orbitals \cite{bauer2014}  or by using mixed classical-quantum dynamics such as the Ehrenfest approach \cite{meyera1979} and the surface hopping method \cite{tully1971,tully1990}. In this paper, we use another class of low-dimensional objects, namely single-particle pilot waves evolved concurrently with Bohmian trajectories to extract all the physical information about the system.

Within the de Broglie-Bohm interpretation of quantum mechanics \cite{durr2012,durr2009,holland1995}, the quantum mechanical wavefunction is a pilot wave that guides the motion of the particle in the physical space. While this interpretation does not alleviate the need for dealing with many-dimensional functions, the prospect of replacing the full many-particle wavefunction by single-particle wavefunctions that guide the Bohmian particles in the physical space was recently explored  \cite{norsen2015}. However, this idea was only applied when the entanglement between the particles could be neglected \cite{norsen2015,oriols2007}, thus ruling out its application to strongly correlated systems. Other approaches to treat many-body wavefunctions with trajectories involve approximations such as the mean-field approximation \cite{christov2017,christov2006} or  the semi-classical approaches \cite{norsen2016,struyve2015} or assume a   wavefunction of a certain form \cite{christov2007}.

In the de Broglie-Bohm interpretation, quantum effects are captured by a so-called quantum potential which together with the classical potential governs the motion of the particle, see, e.g., \cite{lopreore1999}. It has been shown recently  that this term can be computed by modeling quantum phenomena by many interacting classical worlds \cite{hall2014,herrmann2017}.  In this paper, we introduce a novel approach to model quantum phenomena using interacting configurations of quantum particles guided by pilot waves. This approach simulates the multi-particle quantum dynamics in a non-perturbative manner  without neglecting the entanglement or relying on particular assumptions about the underlying quantum state. We apply our approach to study the breathing dynamics of  few-boson systems in a trap with long- and short-range interactions and compute the ground state energy for an exactly solvable system. The proposed approach in its current stage does not supersede established numerical methods nor overcomes the scaling problem of simulating-many body systems, but offers a new way forward that may be further developed into a full-fledged method.

\section*{Results}

Let us illustrate the usage of pilot waves in a 1D system consisting of 2 particles. The coordinates are denoted by $x_1$ and $x_2$, the potential by $V(x_1,x_2)$ and the wavefunction describing the full system is $\Psi(x_1,x_2,t)$. In order to evolve the Bohmian trajectories $X_1(t)$ and $X_2(t)$ for the two particles (we  denote the Bohmian trajectories throughout this paper by uppercase letters), we need to evaluate the pilot waves $\psi_i(x_i,t)$. The pilot waves are the full wavefunction projected on the coordinates of all the particles except one, i.e., $\psi_i(x_i,t)\equiv\Psi(x_1,x_2,t)|_{x_j=X_j(t),\ j\neq i}$; hence, they are also called conditional wavefunctions (CWs) \cite{durr1992}. 
In the absence of gauge fields, the Bohmian velocities are computed in terms of the pilot waves as
\begin{equation}
 \frac{dX_i}{dt}=\left.\frac{\hbar}{m_i}\text{Im}\left\lbrace \frac{\partial_{x_i} \psi_i(x_i,t)}{\psi_i(x_i,t)} \right\rbrace \right|_{x_i=X_i(t)},
 \label{speed}
 \end{equation}
where $m_i$ is the mass of particle $i$.

It is guaranteed that the density of the Bohmian particles evolved by Eq. (\ref{speed}) follows the evolution of the density function as computed by Schrödinger's equation \cite{benseny2014}. In order to evolve the CWs without having to solve the time-dependent Schrödinger equation (TDSE) for $\Psi(x_1,x_2,t)$, we introduce a generalized set of conditional wavefunctions $\psi_i^{n}(x_i,t)$ defined as
\begin{equation}
\psi_i^{n}(x_i,t)\equiv \left. \frac{\partial^n \Psi(x_1,x_2,t) }{\partial x_j^n}\right|_{x_j=X_j(t),\ j\neq i}
\end{equation}
where the pilot waves correspond to $\psi_i^{0}$. 
The equation of motion for  $\psi_i^{n}(x_i,t)$ is given by \cite{norsen2010}
\begin{widetext}
\begin{equation}
i\hbar \frac{\partial\psi_i^n(x_i,t)}{\partial t}=-\frac{\hbar^2}{2 m_i}\frac{\partial^2\psi_i^n(x_i,t)}{\partial x_i^2}+\left.\sum_{k=0}^n\binom{n}{k}\psi_i^{n-k}(x_i,t) \frac{\partial^kV(x_i,x_j)}{\partial x_j^k}\right|_{x_j=X_j(t),\ j\neq i}-\frac{\hbar^2}{2 m_j}\psi_i^{n+2}(x_i,t)+i\hbar \frac{dX_j(t)}{dt}\psi_i^{n+1}(x_i,t) 
\label{eom}
\end{equation}
\end{widetext}
We see from this equation that the pilot waves corresponding to different particles interact indirectly through the last three terms of Eq. (\ref{eom}). In \cite{norsen2015}, a similar equation of motion is derived in terms of nonlocal potentials.

In order to evolve the pilot waves in an exact manner using Eq.  (\ref{eom}) instead of
evolving the multi-dimensional full wavefunction,
 we need to evolve the whole hierarchy of $\{\psi_i^{n}\}$. We illustrate in the Methods section  that  truncating this hierarchy at a finite order $N$ is not an efficient method to obtain the correct dynamics of an entangled system as the truncation errors propagate very quickly to $\psi_i^0$. 
 Is there a way to avoid the errors originating from the truncated orders? It turns out that the answer is yes!  To this end, we  assume an ansatz for the full wavefunction that allows the calculation of the first and second CWs $\psi_i^{1}$ and $\psi_i^{2}$ which we subsequently use to evolve the interacting pilot waves $\psi_i^0$.

{\bf Interacting Pilot Waves.}

The most general form for the wavefunction of a 2-particle system is
\begin{equation}
\Psi(x,y,t)=\sum_{i,j}c_{ij}(t)\phi_i(x)\phi_j(y),
\label{ansatz}
\end{equation}
where $\{\phi_i\} $ is a complete basis for the one-body Hilbert space, also referred to as 'orbitals' later. Let us assume that a finite number of basis states $M$ is sufficient to capture all the important features of the wavefunction. The conditional wavefunctions of the first particle conditioned on the second particle located at $y=Y$ are expressed as (the time variable and the particle index are omitted to simplify the notation) 
\begin{align*}
\psi^{0}(x)\equiv \left.  \Psi(x,y) \right|_{y=Y} =\sum_i a_i \phi_i(x) \\
\psi^{1}(x)\equiv \left. \frac{\partial \Psi(x,y) }{\partial y}\right|_{y=Y} =\sum_i b_i \phi_i(x) \\
\psi^{2}(x)\equiv \left. \frac{\partial^2 \Psi(x,y) }{\partial y^2}\right|_{y=Y} =\sum_i c_i \phi_i(x)
\end{align*}
where $a_i=\sum_{j}c_{ij}\phi_j(Y)$, $b_i=\sum_{j}c_{ij} \frac{\partial  \phi_j(y)} {\partial y}|_{y=Y}$ and $c_i=\sum_{j}c_{ij} \frac{\partial^2  \phi_j(y)} {\partial y^2}|_{y=Y}$. These relations can be written in vector form as:
\begin{equation}
\bm{\vec{a}}(Y)=\bm{C \vec{\phi}}(Y), \bm{\vec{b}}(Y)=\bm{C \vec{\phi}}'(Y) \ \text{and}\ \bm{\vec{c}}(Y)=\bm{C \vec{\phi}}''(Y)  
\label{linear}
\end{equation}
where $\bm{\vec{a}}(Y)=\{a_1(Y),\ a_2(Y),...\}$, $\bm{\vec{\phi}}(Y)=\{\phi_1(Y), \phi_2(Y),...\}$,...etc. 

\begin{figure*}[!t] \setlength{\unitlength}{0.1cm}
\begin{picture}(180,60)
{

\put(0, 0){ \includegraphics[width=\textwidth ]{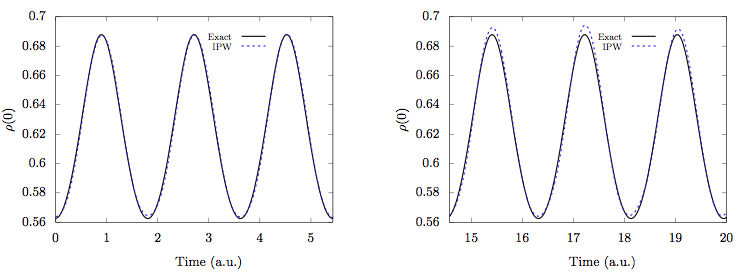} }

}

\end{picture}

\caption{ \label{2bosons}  {\bf The breathing dynamics of two interacting bosons in a harmonic trap.} The two bosons are initially in the ground state of the trap before the harmonic interaction between them is suddenly switched on at $t=0$. The value of the reduced one-body density function $\rho(x)$ at the origin is monitored during the evolution. The strength of the harmonic interaction is the same as the trap strength. We plot $\rho(0)$ computed by the Interacting Pilot Waves method (IPW) using 6 orbitals and by the exact two-body wavefunction (Exact). The IPW results were obtained by averaging over 5000 Bohmian configurations of the two particles. The two panels depict different time slices to illustrate the accuracy in the long-time regime.}

\end{figure*}

The problem of finding $\psi^{1}$ and $\psi^{2}$ boils down to finding the coefficients $b_i$ and $c_i$ constituting the vectors $\bm{\vec{b}}$ and $\bm{\vec{c}}$. This is accomplished by making use of an ensemble of Bohmian pairs of coordinates $\{X,Y\}$ which are selected initially from the one-particle density function $\rho(x)$ at $t=0$ \cite{note}. Each of these pairs is called a configuration. If we can represent both $\bm{\vec{\phi}}'$ and $\bm{\vec{\phi}}''$  for a certain  value of $Y$ as a linear superposition of all $\{\bm{\vec{\phi}}(Y_k)\}$ corresponding to all members of the ensemble, i.e., if  $\bm{\vec{\phi}}'(Y)=\sum_k \alpha_k \bm{\vec{\phi}}(Y_k)$ where $\bm{\vec{\phi}}(Y_k)$ corresponds to the $k^\text{th}$ member of the ensemble and $\bm{\vec{\phi}}''=\sum_k \beta_k \bm{\vec{\phi}}(Y_k)$ then it follows from the linearity in Eq. (\ref{linear})  that $\bm{\vec{b}}(Y)=\sum_k \alpha_k \bm{\vec{a}}(Y_k)$ and $\bm{\vec{c}}(Y)=\sum_k \beta_k \bm{\vec{a}}(Y_k)$. Finding the values of $\alpha_k$ and $\beta_k$ is equivalent to solving a system of linear equations. In this way, we can obtain $\psi^1$, $\psi^2$ without ever constructing the coefficient matrix $\bm{C}$.   

\begin{figure*}[!t] \setlength{\unitlength}{0.1cm}
\begin{picture}(180,130)
{

\put(0, 0){ \includegraphics[width=\textwidth ]{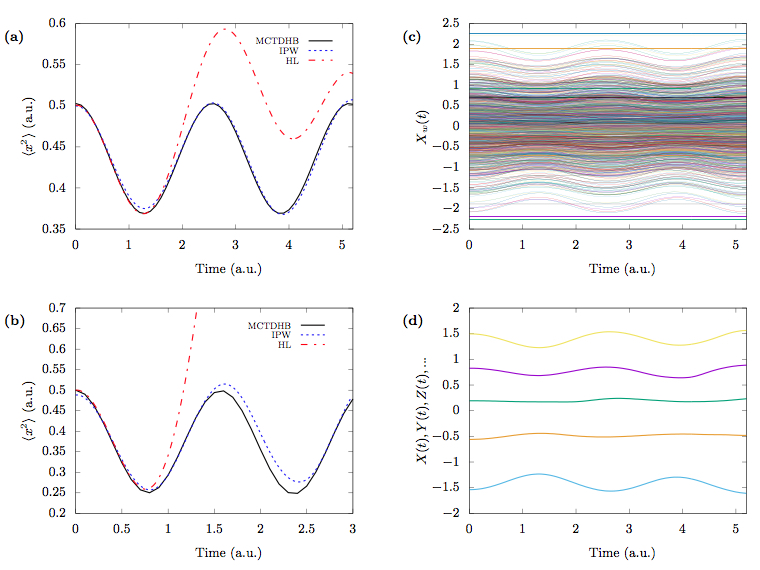} }

}

\end{picture}

\caption{ \label{5bosons}  {\bf Breathing dynamics of 5  bosons and 3 bosons using IPW compared with numerically exact dynamics.}  (a,b) 5 bosons and 3 bosons are initially in the ground state of a harmonic trap before a harmonic interaction of strength $k_i=0.1$ and $k_i=1$ respectively is switched on at $t=0$. The breathing dynamics is computed by the Interacting Pilot Wave (IPW) method for 5 and 3 bosons using 3 and 4 orbitals  respectively  and compared with the numerically exact dynamics computed by  MCTDHB  and with the Hermitian limit (HL) solution of Eq. (\ref{eqn}) (see text).  In (c) we depict the time evolution for an ensemble of 1000 Bohmian trajectories representing the first particle, and in (d) the 5 trajectories corresponding to  a single configuration for the case (a). }

\end{figure*}


It should be noted that $\psi^1$ and $\psi^2$ can be determined without expressing $\psi^0$ in terms of a basis at all, since $\{\alpha_k\}$ and $\{\beta_k\}$ depend only on the amplitudes of any complete basis at the location of the Bohmian particles. After  $\{\alpha_k\}$ and $\{\beta_k\}$ are obtained, we can express $\psi^1$ and $\psi^2$ as 
\begin{align*}
\psi^{1}(x) =\sum_k \alpha_k \psi^{0}_k(x;Y_k) \\
\psi^{2}(x) =\sum_k \beta_k \psi^{0}_k(x;Y_k)
\end{align*}

 With the CWs at our disposal,  we use the equation of motion (\ref{eom}) for $n=0$ to evolve the  ensemble of CWs for all Bohmian particles as described in the Methods section. We call this scheme Interacting Pilot  Waves (IPW).

 Some observables can be computed by averaging over the ensemble of Bohmian configurations $\{X,Y\}$ such as $\langle x^2 \rangle$. Since we have access to the CWs, we can devise a more accurate method that approximates the exact expression of the expectation value of an operator $\hat{A}$, $\langle \hat{A} \rangle =\int \Psi^*(x,y) \hat{A} \Psi(x,y) dxdy$ by performing the integral over one variable as a Riemann sum over its Bohmian coordinates, i.e., 
 \begin{equation}
\langle \hat{A} \rangle \approx \sum_w \Delta_w \int \psi_w^*(x) \hat{A}_w \psi_w(x) dx,
\label{expectaion}
 \end{equation} 
where $\psi_w(x)$ is the conditional wavefunction  of the first particle conditioned on the coordinate of the second particle belonging to the $w^{ \text{th}}$ configuration of the ensemble, $\Delta_w$ is the distance between adjacent values of $Y$ at the $w^{ \text{th}}$ configuration and $\hat{A}_w$ is the operator $\hat{A}$ conditioned on $Y_{w}$. For two-body operators such as the interaction potential $V(x,y)$, $\hat{A}_w(x)$ is given by $V(x;Y_w)$.
   Similarly, the reduced one-body density  can be approximated as $\rho(x) \approx \sum_w \Delta_w \psi_w^*(x) \psi_w(x) $.

Let us apply this method to study the breathing dynamics of two bosons in a harmonic trap, $V(x)=\frac{1}{2} k_t x^2$. The bosons are initially condensed in the ground state of the trap, and start a breathing motion when a harmonic interaction $V(x,y)=\frac{1}{2} k_i (x-y)^2$ is suddenly switched on. A finite fixed set of orbitals are taken to be the lowest set of eigenfunctions of the one-body problem with the effective potential felt by one particle due to the other one, namely $V_{\text{eff}}(x)=\frac{1}{2} k_t x^2+ \int \rho(y) V(x,y) dy $.  We illustrate in Fig. \ref{2bosons} the behavior of $\rho(0)$ for $k_t=k_i=1$ computed by the IPW method with 6 orbitals compared with the exact dynamics (atomic units are used in the rest of this paper).

{\bf Generalization to many-particle systems.}

Generalizing our algorithm to a many-particle problem consisting of $N_B$ bosons is straightforward. Let us denote the coordinates of the particles by $x, y,z,\cdots$ etc., while, as before, we denote the Bohmian coordinates by upper case letters. A single configuration of Bohmian walkers is denoted by $\left( X,Y,Z, \cdots \right)$.

Let us denote the conditional wavfunctions by
\begin{align*}
\psi^0(x;Y,Z,\cdots) \equiv \left. \Psi(x,y,z,\cdots)\right|_{y=Y,z=Z,\cdots}\\
\psi^1(x;Y',Z,\cdots)\equiv \left. \frac{\partial \Psi(x,y,z,\cdots)}{\partial y}\right|_{y=Y,z=Z,\cdots}\\
\psi^1(z;X,Y',\cdots) \equiv \left. \frac{\partial \Psi(x,y,z,\cdots)}{\partial y} \right|_{x=X,y=Y,\cdots},
\end{align*}
and so on. The equation of motion for $\psi^0(x;Y,Z,\cdots)$ is a simple generalization of Eq. (\ref{eom}) for the case of many particles:
\begin{align}
 i \hbar \partial_t \psi^0(x;Y,Z,\cdots;t)=\nonumber\\
 \left(-\frac{\hbar^2}{2 m} \nabla_x^2+V(x;Y,Z,\cdots)\right)\psi^0(x;Y,Z,\cdots;t)\nonumber\\
+i \hbar\frac{dY}{dt}\psi^1(x;Y',Z,\cdots;t)-\frac{\hbar^2}{2 m}\psi^2(x;Y'',Z,\cdots;t)\nonumber\\
+i \hbar\frac{dZ}{dt}\psi^1(x;Y,Z',\cdots;t)-\frac{\hbar^2}{2 m}\psi^2(x;Y,Z'',\cdots;t)+\nonumber\\
+\cdots \text{etc.} 
\label{eqn}
\end{align}
Similar equations can be written for all the  CWs corresponding to  all particles in every configuration. A generic ansatz for the many-body wavefunction similar to Eq. (\ref{ansatz}) reads
\begin{equation}
\Psi(x,y,z,...,t)=\sum_{i,j,k}c_{ijk...}(t)\phi_i(x)\phi_j(y)\phi_k(z)z... 
\label{ansatz2}
\end{equation}

In order to compute $\psi^1(x;Y',Z,\cdots;t)$ from $\psi^0_w(x;Y_w,Z_w,\cdots;t)$ belonging to all configurations, we need to express the tensor $[\phi_j'(Y)\phi_k(Z)...]$ as a linear superpositions of all the $[\phi_j(Y_w)\phi_k(Z_w)...]$ tensors belonging to all configurations; i.e., $[\phi_j'(Y)\phi_k(Z)...]=\sum_w \alpha_w  [\phi_j(Y_w)\phi_k(Z_w)...]$. This can be done with the existing  numerical techniques by rearranging all the $M^{N_B-1}$ terms of the tensors, where $M$ is the number of orbitals, in vector forms and solving  a linear system of equations. Since the size of the vectors now becomes exponentially bigger as the number of particles becomes larger, the bottleneck of this method would be to take a sufficiently large number of configurations that ensures having a complete linear system. Therefore, an immediate room for improvement here would be to find smart tactics to overcome this problem.

{\bf  Computing observables from conditional wavefunctions  }
    
    In order to compute the expectation value of an operator, we can treat the collection of normalized CWs belonging to all the particles as if they describe normal single particle wavefunctions. $\langle \hat{A}_x \rangle$ can then be computed as $\langle \hat{A_x} \rangle \approx \frac{1}{N_w} \sum_w \int \tilde{\psi}_w^*(x) \hat{A}_x \tilde{\psi}_w(x) dx$, where $\tilde{\psi}_w(x)$ is the normalized CW of particle $x$ belonging to the $w^\text{th}$ configuration and $N_w$ is the number of configurations. If $\hat{A}$ is a two-body operator such as the interaction potential between two particles, we first compute a mean-field operator, and then treat it as a one-body operator. For example, the mean-field interaction potential felt by one particle is computed as $\tilde{V}(x)  \approx \frac{1}{N_w} \sum_w \int \tilde{\psi}_w^*(y) V(x-y) \tilde{\psi}_w(y) dy$. The expectation value of $\tilde{V}(x)$ is then computed as a one-body operator. From the normalized collection of all the CWs, we can also get an approximation for the reduced density matrix of one particle $\rho(x',x) \approx \frac{1}{N_w} \sum_w\tilde{ \psi}_w(x')  \tilde{\psi}_w^*(x) $.   This matrix can be used to compute a set of natural orbitals in terms of the finite basis set used in the postulated ansatz as the eigenstates of $\rho(x',x)$. 

{\bf  Many particles in a harmonic trap  }     
     
    Let us apply this generalization to the dynamics of 3 bosons and 5 bosons in a harmonic trap $V(x)=\frac{1}{2}x^2$ for two cases of interparticle interactions: long-range attractive harmonic interaction  and short-range repulsive interaction. As in the two-particle case, all the bosons initially reside in the ground state of the harmonic trap before the interaction is suddenly switched on at $t=0$. We study the breathing dynamics by computing $\langle x^2 \rangle$ as a function of time.

For a harmonic interaction of the form $V(x,y)=\frac{1}{2}k_i(x-y)^2$ we consider two cases of 5 bosons with weak interactions  ($k_i=0.1$) and 3 bosons with strong  interaction ($k_i=1$)  and we use 3 and 4 orbitals in the two cases, respectively. In both cases we compare the results with the numerically exact simulation using multiconfigurational time-dependent Hartree method for bosons (MCTDHB) \cite{alon2008,lode2016,fasshauer2016,mctdh-x, mctdh-b} and with the Hermitian limit (HL) of Eq. (\ref{eqn}) (also referred to as small entanglement approximation) where all the non-hermitian terms in Eq. (\ref{eqn}) are dropped out. The Hermitian limit is equivalent to the time-dependent quantum Monte-Carlo (TDQMC) of Ref. \cite{christov2007} which does not take entanglement into consideration. It was also employed recently in \cite{albareda2014,albareda2015} in order to devise an approximate solution for electron-nuclear dynamics in molecular systems.

In Fig. \ref{5bosons}, we show the results of computing $\langle x^2 \rangle$ by averaging over the Bohmian coordinates of all the  particles using a single ensemble containing 1000 configurations.  We notice that the IPW method is more accurate in the weak interaction regime than in the strong interaction regime. The Bohmian trajectories of the first particle in all  configurations are shown in Fig. \ref{5bosons}(c) for $k_i=0.1$ while the Bohmian trajectories for all the 5 particles in a single configuration are shown in Fig. \ref{5bosons}(d). The few constant trajectories appearing in Fig. \ref{5bosons}(c) correspond to the cases where we manually set the Bohmian velocities to be zero when the denominator in Eq. (\ref{speed})  is below a certain threshold.


\begin{figure}[] \setlength{\unitlength}{0.1cm}
\begin{picture}(90,60)
{
\put(0, 0){\includegraphics[width=\columnwidth]{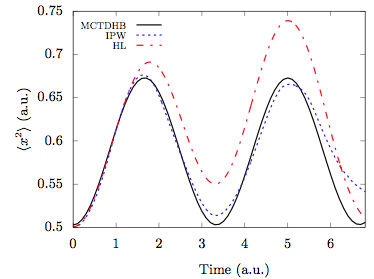} }
}
\end{picture}

\caption{ \label{5bosons-hard} {\bf Breathing dynamics of 5  bosons with short-range repulsive interactions.} Same as in Fig. \ref{5bosons} with inter-particle gaussian interactions $V(x-y)=k_i/\sqrt{2\pi\sigma^2}\times\text{e}^{\frac{-(x-y)^2}{2\sigma^2}}$ with $k_i=0.1$, $\sigma=0.25$. Three orbitals are used in the IPW calculation.}

\end{figure}

 In Fig. \ref{5bosons-hard}, we plot $\langle x^2 \rangle$ after switching on a gaussian interaction $V(x-y)=k_i/\sqrt{2\pi\sigma^2}\times\text{e}^{\frac{-(x-y)^2}{2\sigma^2}}$ with $k_i=0.1$, $\sigma=0.25$ and compare the results with MCTDHB simulation and the HL of Eq. (\ref{eqn}). In this calculation, we compute $\langle x^2 \rangle$ from the expectation value of ${x}^2$ using the conditional wavefunctions rather than from the Bohmian trajectories.\\ 

\begin{figure}[] \setlength{\unitlength}{0.1cm}
\begin{picture}(90,60)
{
\put(0, 0){ \includegraphics[width=\columnwidth ]{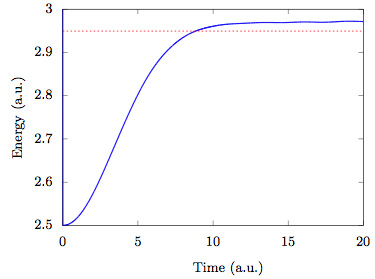} }
}

\end{picture}

\caption{ \label{5b-En}  {\bf Ground state energy computation using IPW.  } Ground state energy for a system of 5 interacting bosons in a harmonic trap with harmonic interaction strength $k_i=0.1$  computed by the IPW method (solid line) using 3 orbitals  and compared with the exact ground state energy (dashed). The system is initialized in the ground state of the trap before the interaction is switched on adiabatically as $k_i=0.1\times(1-e^{-0.02t^2})$ . The energy is computed with respect to the instantaneous value of $k_i$.}

\end{figure}

     
 In order to compute the ground state energy for an interacting system of particles using the IPW scheme, we initialize the CWs and the Bohmian trajectories in the ground state of the noninteracting Hamiltonian. Afterwards, we switch on the interaction adiabatically. According to the adiabatic theory \cite{Born1928}, the system remains in the ground state of the instantaneous Hamiltonian. In Fig. \ref{5b-En}, we plot the evolution of the energy of the instantaneous Hamiltonian of a 5-particle system as we switch on the harmonic interaction $V(x,y)=\frac{1}{2}k_i(x-y)^2$  adiabatically and compare it with the exact ground state energy $E_0=\frac{N_B-1}{2}\sqrt{1+k_iN_B}+0.5$ \cite{yan2003} for $k_i=0.1$.   The ground state energy computed by MCTDHB \cite{lode2012} is more accurate than the IPW calculation for the same number of orbitals by several significant digits. Perhaps a better method to compute ground state energy is to propagate  Eq. (\ref{eqn}) in complex time, while evolving the Bohmian trajectories in real time \cite{christov2007}.  The optimal relation between the real and complex time  evolution constitutes an interesting topic of research. \\

{\bf Discussion }

We have presented a promising approach to analyze the dynamic and static properties of systems consisting of several bosons by evolving a system of nonunitary equations that goes beyond the small entanglement approximation and the mean-field approximation.  
Our method builds on the formal expansion (Eq. \ref{eom}), but as we find that the truncation of this set of equations quickly leads to errors, we introduce and apply a truncation-free method that provides the lowest order pilot waves in a self-consistent manner.

The accuracy of this new approach is confirmed but also outperformed by the state-of-the-art MCTDHB algorithm. In the MCTDHB method, increasing the  number of orbitals ($M$) is confronted with the exponentially large number of configurations of permanents that needs to be taken into account. We have a similar scaling problem in our approach; the number of configurations of Bohmian particles has to be larger than $M^{N_B-1}$ in order to avoid having an undetermined linear system of equations when solving for $\alpha_w$. So, the complexity of our approach still increases exponentially with the number of particles.        It is worth mentioning also that MCTDHB is much faster than our algorithm. While a typical result in the previous figures takes a few hours to compute, it takes much less time by the well developed MCTDHB.    

Improvements on our method may come from: (i)  strategies that minimize the number of configurations, and hence the computational power, required to evolve the pilot waves without having an underdetermined system of equations, and (ii) optimal choice of the basis functions (possibly an adaptive set of orbitals) that lead to the most compact representation of the full wavefunction, and hence the smallest number of orbitals to capture the dynamics of the many-particle wavefunction.

It is still an open question whether the non-Hermitian terms in the equations of motion can be replaced by an effective entanglement potential that makes the equations unitary and at the same time captures the entanglement in the system.
For systems consisting of many particles, i.e., $N_B>20$, the entanglement of the ground state is so small \cite{walmsley2017} that even the Hermitian limit \cite{christov2007} can be efficient for simulating the dynamics involving a small number of excited states.

In principle, generalizing the IPW approach to fermionic systems is straightforward, as long as we choose the initial state with the proper symmetry requirements. However, for fermions two problems arise. First, due to the Pauli exclusion principle, we need a large number of orbitals to describe a fermionic state and therefore, the number of fermions that can be analyzed is  small compared to bosons. Second, the conditional wavefunctions for fermions will have nodes, that complicate computing the velocity of the Bohmian walkers around those nodes. Since the node problem is a well known problem for simulating quantum dynamics with Bohmian trajectories \cite{wyatt2006},  the methods developed in this regard in the literature  \cite{goldfarb2007,Bittner2012} may benefit the solution of this problem.

As a final comment, we note that in order to describe entangled dynamics, we need to consider many interacting configurations of Bohmian particles in the present approach. A similar situation arises in \cite{hall2014,herrmann2017} where the concept of interacting classical worlds was introduced. 
This similarity between the two approaches may be worth further attention in discussions of the foundations of quantum mechanics.”

\clearpage
\begin{widetext}

{\bf \vspace{4mm} \Large Methods \vspace{4mm} }

{\bf  Derivation of Equation (\ref{eom}) }

Let us derive Eq. (\ref{eom}) with respect to $\psi_1^n(x_1,t)$. Since $\psi_1^n(x_1,t) \equiv \frac{\partial^n \Psi(x_1,x_2,t) }{\partial x_2^n}|_{x_2=X_2(t)}$, we find by the chain rule that 
\begin{equation}
 \frac{\partial\psi_1^n(x_1,t)}{\partial t}=\frac{\partial}{\partial t} \left. \frac{\partial^n\Psi(x_1,x_2)}{\partial x_2^n} \right|_{x_2=X_2(t)}+\frac{dX_2(t)}{dt}\left. \frac{\partial^n\Psi(x_1,x_2)}{\partial x_2^n} \right|_{x_2=X_2(t)}.
 \label{eom5}
 \end{equation} 
 By exchanging the time and spatial derivatives in the first term on the R.H.S. and using the TDSE:
\[
i\hbar\frac{\partial \Psi(x_1,x_2)}{\partial t}=\left(-\frac{\hbar^2\nabla_1^2}{2m_1}-\frac{\hbar^2\nabla_2^2}{2m_2}+V(x_1,x_2)\right)\Psi(x_1,x_2),
\]
we obtain after substituting back in Eq. (\ref{eom5})

\begin{equation}
i\hbar \frac{\partial\psi_1^n(x_1,t)}{\partial t}=-\frac{\hbar^2}{2 m_1}\frac{\partial^2\psi_1^n(x_1,t)}{\partial x_1^2}+\frac{\partial^n}{\partial x_2^n}\left.\left(V(x_1,x_2) \Psi(x_1,x_2)\right)\right|_{x_2=X_2(t)}
-\frac{\hbar^2}{2 m_2}\psi_1^{n+2}(x_1,t)+i\hbar \frac{dX_2(t)}{dt}\psi_1^{n+1}(x_1,t). 
\label{eom6}
\end{equation}

By applying the chain rule to the second term on the R.H.S. we obtain  
$\frac{\partial^n}{\partial x_2^n}\left.\left(V(x_1,x_2) \Psi(x_1,x_2)\right)\right|_{x_2=X_2(t)}=\sum_{k=0}^n\binom{n}{k}\psi_1^{n-k}(x_1,t) \frac{\partial^kV(x_1,x_2)}{\partial x_2^k}|_{x_2=X_2(t)}$ which after substituting in Eq. (\ref{eom6}) recovers Eq.  (\ref{eom}).\\

{\bf  Evolving a hierarchy of pilot waves }

Let us illustrate the inefficiency of evolving a truncated hierarchy of $\psi_i^{n}$  using Eq. (\ref{eom}) in order to compute the dynamics of an entangled system. 
We consider the entangled dynamics of two particles of masses $m_1=1$ and $m_2=100$ subject to the harmonic potential
$V(x_1,x_2)=\frac{1}{2} k x_1^2+\frac{1}{2} k x_2^2$ with $k=0.1$. Let us take the initial  state
to be the entangled ground state of the Hamiltonian with the potential function
$V(x_1,x_2)=\frac{1}{2} k_1 x_1^2+\frac{1}{2} k_2 x_2^2+\frac{1}{2} k_3 (x_1-x_2)^2$ with $k_1=k_2=0.1$, $k_3=1.0$ and the masses of the particle  $m_1=1$ and  $m_2=2$. This is an entangled state. We evolve the Bohmian trajectories for  the initial conditions $X_1=1$, $X_2=2$. We first truncate the hierarchy at $N=0$, thus making Eq. (\ref{eom}) unitary. This case corresponds to  the Hermitian limit, i.e., noninterating pilot waves. Fig. \ref{cwa1} shows that
the Bohmian trajectory evolved by the corresponding pilot wave deviates from the trajectory computed from the exact pilot wave already at half a cycle of the oscillatory motion.  Increasing the depth of the hierarchy to $N = 7$ only extends the range of accurate dynamics for another cycle.

In this example, we see clearly that although the particles are non-interacting, we need to account for the interaction between the pilot waves of the two particles correctly through the higher order CWs  even when the ratio of the particles' masses is 1:100. Otherwise, the errors originating from truncating the hierarchy of $\{\psi_i^{n}\}$ propagate very fast to  $\psi_i^{0}$. Increasing $N$ beyond $N\approx 10$ will not help in prolonging the range of accuracy because of the numerical errors in the calculation of the higher-order derivatives of the wavefunction.

  Since the errors afflict the pilot waves through the last two terms in Eq. (\ref{eom}), this method of evolving the pilot waves is most suitable when we are interested in the dynamics of a very light particle interacting with a much heavier one over a very short time scale. In this case, we can omit the last two terms for the heavy particle while retaining them for the light particle, i.e., do a semiclassical approximation \cite{struyve2015} for one particle only. \\

\begin{figure*}[] \setlength{\unitlength}{0.1cm}
\begin{picture}(180,60)
{

\put(0,0){ \includegraphics[ width=\textwidth]{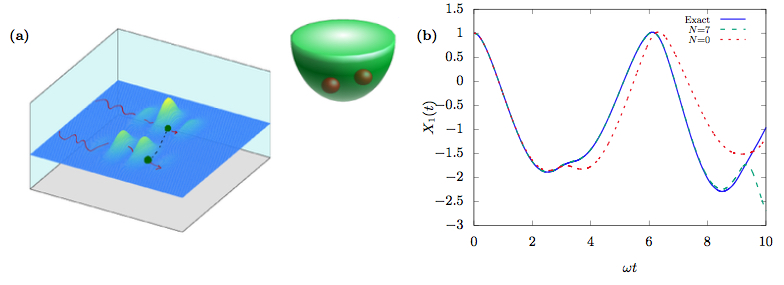}  }

}
\end{picture}
\caption{\label{cwa1} 
{\bf Entangled quantum dynamics of  two particles in a harmonic trap.}  (a) A cartoon representing the pilot waves guiding Bohmian particles moving in 2D. Although the two particles do not interact, their entanglement implies a coupling between the two pilot waves guiding the two Bohmian trajectories (the dashed line).  (b) The trajectories of the light particle computed from the exact pilot waves (solid blue) and using pilot waves evolved using Eq. (\ref{eom}) (dashed). The case for $N=7$ corresponds to interacting pilot waves evolved with the help of a hierarchy of conditional wavefunctions $\{\psi_1^{n},\psi_2^{n}\}$ up to $n=7$ (see Eq.(\ref{eom})), while  $N=0$ corresponds to noninterating pilot waves (the Hermitian limit). We notice that the former case is more accurate than the latter. The two particles have mass ratio 1:100. $\omega$ is the trap frequency of the light particle.}

\end{figure*}


{\bf  Interacting Pilot Waves for two particles }

For the two-particle IPW simulations in Fig. \ref{2bosons} we express all the CWs in terms of the basis set $\{\phi_i\} $, and evolve the expansion coefficients $\{a_i\}$.  Equation (\ref{eom}) is then expressed as
 
\begin{equation}
 i\hbar \sum_i \dot{a}_i(t) \phi_i(x) =-\frac{\hbar^2}{2 m} \sum_i a_i(t) \frac{\partial^2\phi_i(x)}{\partial x^2}+V(x,Y)\sum_i a_i(t) \phi_i(x)  -\frac{\hbar^2}{2 m}  \sum_i c_i(t) \phi_i(x) +i\hbar \frac{dY}{dt}\sum_i b_i(t) \phi_i(x).
 \label{eom2}
 \end{equation}

 By taking the inner product with each of the orbitals $\{\phi_i(x)\}$ we  obtain the time derivative of the expansion coefficients $\{\dot{a}_i\}$. This system of  equations is then solved using a fourth-order Runge-Kutta method. \\

{\bf  Propagating the  conditional wavefunctions }
\\
Each of equations (\ref{eom}) and (\ref{eqn}) represents a system of coupled nonlinear and nonunitary differential equations that  can be cast in the form 

\begin{equation}
i\frac{\partial\psi(x,t)}{\partial t}=H\psi(x,t)+W(x,t),
\label{eom3}
\end{equation}
where the first term on the RHS represents the unitary part of the equation and $W(x,t)$ represents the nonunitary part which is a function of all other CWs. If, e.g., $H$ is a constant Hamiltonian, a general solution for this equation takes the form: 
\[
\psi(x,t)= e^{-iHt}\left [ \int_0^t e^{iHt'}W(x,t')dt'+\psi(x,t_0) \right ]
\]
In order to propagate $\psi(x,t)$  for a single time step from $t=0$ to $t=\delta t$ using this solution, both the operator $e^{-iH\delta t}$ and $e^{-iHt'}$ are performed using a split-operator method \cite{kosloff1988}. The integral is performed using the trapezoidal rule $ \int_0^{\delta t} e^{iHt'}W(x,t')dt' \sim \frac{1}{2} \left [  e^{iH\delta t}W(x,\delta t)+W(x,0) \right ] $.\\

\vspace{5mm}

{\bf  Miscellaneous numerical techniques }
\begin{itemize}

\item We use the split operator technique [36] in order to propagate the exact two-particle wavefunction in imaginary time (to generate the ground state) and in real time (as in Fig. \ref{2bosons}) or to generate the exact Bohmian trajectories (as in Fig. \ref{cwa1}). 
\item The evaluation of the conditional wavefunctions at the  position of the Bohmian particles was performed by FFT-based interpolation.
\item The spatial derivatives of the wavefunction to compute $\{\psi_1^{n},\psi_2^{n}\}$ are computed using the Fast Fourier Transform with the FFTW package.
\item Solving the linear system of equations in Eq. (\ref{linear}) was performed by the  LAPACK routine {\texttt  gelsd()} which uses  singular value decomposition and a divide and conquer method to compute the minimum-norm solution to a linear least squares problem \cite{mkl}.
\end{itemize} 
\end{widetext}

\clearpage
{\bf Acknowledgments }
T. A. Elsayed thanks D. A. Deckert, H. Miyagi, A. I. Streltsov, L. F. Buchmann, C. Leveque and Andreas Caranti for brief discussions, and I. Christov for his comments on an earlier version of the manuscript. The authors acknowledge  financial support by Villum Foundation. Part of this paper was written while T. A. Elsayed was visiting Hunter College of the City University of New York.

{\bf  Author contributions  }
TAE conceived the research idea, designed the IPW algorithm for two and many particles and implemented it on the computer. LBM and KM provided helpful insights and  feedback during the development of the method. All authors participated in the data analysis and the troubleshooting. The manuscript was written by TAE with extensive feedback from LBM and KM.

{\bf  Competing interests  }
The authors declare no competing interests.

\bibliographystyle{naturemag}

\bibliography{cwa_abrv}

\begin{thebibliography}{10}
\bibitem{note}
 \bibinfo{note}{For a system of bosons, the Bohmian coordinate of each boson is
  picked according to the single particle density $\rho(x)$ which corresponds
  to the diagonal function of the two body density matrix $\rho(x,x’)$,
  where, for a two-body problem, $\rho(x,x’)=\int
  \psi(x,y)\psi^*(x’,y’)dydy’$.}
\end{thebibliography}

\end{document}